\documentclass[11pt]{article}
\usepackage{Blois,epsfig}

\def\Journal#1#2#3#4{{#1} {\bf #2}, #3 (#4)}

\def\NCA{\em Nuovo Cimento}

\def\NPB{{\em Nucl. Phys.} B}
\def\PLB{{\em Phys. Lett.}  B}

\def\be{\begin{equation}}
\def\ee{\end{equation}}
\def\bea{\begin{eqnarray}}
\def\eea{\end{eqnarray}}

\begin{document}
\vspace*{2cm}

\vspace*{2cm}
\title{Calculation of the real part of the nuclear amplitude \\
 at high s and small t from the Coulomb amplitude }

\author{ P. Gauron, B. Nicolescu \\
{\it Theory Group,  Laboratoire de Physique Nucl\'eaire  et des Hautes  \'Energies
(LPNHE)\footnote{Unit\'e  de Recherche des Universit\'es
  Paris 6 et Paris 7, Associ\'ee au CNRS},
 CNRS and Universit\'e Pierre et Marie Curie, Paris\\
{\small e-mail: \texttt{gauron@in2p3.fr} } \\
     {\small e-mail: \texttt{nicolesc@lpnhep.in2p3.fr } \hspace{20cm}} }\\
      \phantom{.}  \hspace{20cm} \\
                            O.V. Selyugin }

\address{Bogoliubov Laboratory of Theoretical Physics, Joint Institute for Nuclear Reserch, Dubna  \\
    {\small e-mail: \texttt{selugin@theor.jinr.ru}}
}
 \maketitle\abstracts{
A new method for the determination of the real part of  the elastic scattering
amplitude  is examined for high energy proton-proton at small momentum transfer. This
method allows us to decrease the number of  model assumptions,  to obtain the real part
in a narrow region  of momentum transfer and to test  different models.
 The possible non-exponential behavior of the real part was found on the base of
  the analysis of  the ISR experimental data.}


\newpage

Numerous  discussions  of the $\rho$-parameter  measured by the UA4 \cite{ua4} and
UA4/2 \cite{ua42} Collaborations in $p\bar p$ scattering at $\sqrt{s}=541$ GeV have
revealed the ambiguity in the definition of this semi-theoretical parameter. As a
result, it has been shown that one has some trouble in extracting from experiment the
total cross sections and the value of the forward ($t=0$) real part of the scattering
amplitudes \cite{selpl,gns97}.

The standard procedure to extract the magnitude of the real part of the hadron elastic
scattering includes a fit to the experimental data by minimizing the $\chi^2$ function.
We assume, as usual, that at high energies and small angles the one-flip and
double-flip amplitudes are small with respect to the spin-nonflip ones and that the
hadronic contributions to $\Phi_1$ and $\Phi_3$ are the same, as are the
electromagnetic ones. Therefore the scattering amplitude can be written as:
\begin{equation}
\label{sup} F (s,t) = F_{N}+F_{C} \ \exp( i \alpha \varphi).
\end{equation}

In the standard fitting procedure, one neglects the $\alpha^2$ term and the
differential cross section has the form:
\begin{eqnarray}
d\sigma/dt = \pi [ (F_{C} (t))^2+ (\rho(s,t)^2 + 1) (Im F_{N}(s,t))^{2})
+ 2 (\rho(s,t)+ \alpha \varphi(t)) F_{C}(t) Im F_{N}(s,t)], \label{ds2}
\end{eqnarray}
where $F_{C}(t) = \mp\ 2 \alpha G^{2}(t)/|t|$ is the Coulomb amplitude (the upper sign
is for $pp$, the lower sign is for $p\bar p$) and $G^2(t)$ is  the  proton
electromagnetic form factor squared; $ReF_N(s,t)$ and $ImF_N(s,t)$ are the real and
imaginary parts of the hadron amplitude; $\rho(s,t) = Re F_N(s,t) / Im F_N(s,t)$. The
formula (\ref{ds2}) is used for the fit of experimental  data in getting hadron
amplitudes and the Coulomb-hadron phase in order to obtain the value of $\rho(s,t)$.

Let us note two points concerning the familiar exponential forms of $ReF_N(s,t)$ and
$ImF_N(s,t)$ used by experimentalists. First, for simplicity reasons, one makes the
assumption that the slope of imaginary part  of the scattering amplitude is equal to
the slope of its real part in  the examined range of momentum transfer, and, for the
best fit, one should take the interval of momentum transfer sufficiently large. Second,
the magnitude of $\rho(s,t)$ thus obtained  corresponds to the whole interval of
momentum transfer .

We define the imaginary part of the scattering amplitude {\it via} the usual
exponential approximation in the small $t$-region
\begin{equation}
\label{im} Im F_{N}(s,t) = \sigma_{tot}/(0.389 \cdot 4 \pi) \exp(B t/2),
\end{equation}
 where 0.389 is the usual converting dimensional factor for expressing $\sigma_{\mbox{tot}}$ in mb.

Let us define the sum of the real parts of the hadron and Coulomb amplitudes as
$\sqrt{\Delta_{R}}$, so we can write:
\begin{equation}
\label{Del}
 \Delta^{th}_{R}(s,t_i) =  [ Re F_{N}(s,t_i)+ F_{C}(t_i)]^2\geq 0\ .
\end{equation}
Using  the experimental data on the differential cross sections we obtain:
\begin{equation}
\label{Del2} \Delta^{exp}_{R}(s,t_i)  =   (1/\pi) \  d\sigma^{exp}/ dt(t_{i}) - (
\alpha \varphi F_{C}(t_i)+Im F_{N}(s,t_i))^2\ .
\end{equation}
Let us note that the real part of the Coulomb $pp$ scattering
amplitude is negative and exceeds the size of $F_N^{pp} (s,t)$ at $t \rightarrow 0$,
but has a large slope. As the real part of the hadron amplitude is known as being
positive at relatively high (ISR) energies, it is obvious that  $\Delta_R^{th}$ must go
through zero at some value $t=t_{min}^{pp}$ and therefore $\Delta_R^{exp}$ must have a
minimum at the same value $t=t_{min}^{pp}$ at which we have the remarkable equality
\begin{equation}
\label{ReFN} ReF_N^{pp}(t_{min}^{pp})=-F_C^{pp}(t_{min}^{pp})\ .
\end{equation}
The minimum of $\Delta_R^{exp}$ corresponds to a \textit{zero} in $\Delta_R^{th}$ at
some fixed $s$
\begin{equation}
\label{DthR} \Delta_R^{th}(s,t_{min}^{pp})=0\ .
\end{equation}

Let us finally note that our method gives a powerful test for the exponential forms of
$ReF_N^{pp}(s,t)$ and $ImF_N^{pp}(s,t)$. Namely, in the case of these exponential
forms, we have
\begin{equation}
\label{rhost}
\rho^{pp}(s,t)=\frac{ReF_N^{pp}(s,t)}{ImF_N^{pp}(s,t)}=\rho^{pp}(s,0)=\mbox{ const
}=\rho^{pp}(s,t_{min})\ .
\end{equation}
However our method gives the possibility to extract $\rho^{pp}(s,t_{min}^{pp})$
\textit{without} assuming the exponential form for $ReF_N^{pp}(s,t)$, from
eqs.~(\ref{im}) and (\ref{ReFN}). If this numerical value of
$\rho^{pp}(s,t_{min}^{pp})$ is significantly different from the value $\rho^{pp}(s,0)$
extracted by a given experiment, this means that the exponential form of
$ReF_N^{pp}(s,t)$, supposed to be identical to that of $ImF_N^{pp}(s,t)$, is doubtful.

 The problem here is that we extract a small
quantity - the real part of the hadron elastic amplitude - affected by large errors. In
order to minimize these errors we need a very high-precision experiment. The only $pp$
data we did find in literature, satisfying our criterium, are those at $\sqrt{s}=52.8$
GeV \cite{528}.
 In Fig. 1a we plot $\Delta_R^{exp}(s,t_i)$ as given by eqs. (\ref{Del2})
, with $\sigma_T^{pp}=42.38$ mb and $B^{pp}=12.87 \mbox{ (GeV)}^{-2}$
\cite{528}. The error bars of the $\Delta_R^{exp}$ points are calculated from the
errors bars of $d\sigma^{exp}/dt$ points.
 We also plot on the same figure
$\Delta_R^{th}(s,t_i)$ as given by Eq. (\ref{Del}), where
\begin{equation}
\label{new15} ReF_N^{pp}(s,t)=(\rho^{pp}\cdot\sigma_{tot}^{pp})/(0.389\cdot
4\pi)\exp(B^{pp}t/2),
\end{equation}
with $\rho^{pp}=0.077$.
 We see from Fig. 1a that there is a clear disagreement between
$\Delta_R^{th}(s,t_i)$ and $\Delta_R^{exp}(s,t_i)$ in the region
$ 0.03<-t<0.06 \mbox{ GeV}^2\ $.
Namely, $\Delta_R^{th}(s,t_i)$ goes through zero at $-t\simeq 0.024  \mbox{ GeV}^2$
while $\Delta_R^{exp}(s,t_i)$ goes through a minimum at a very different value of $t$.
Moreover, the values of the two quantities are very different
 in the region $ -t > 0.03 \mbox{ GeV}^2\ $.
 In fact the entire shape of $\Delta_R^{th}$ in the above region of $t$
is not consistent with the shape of $\Delta_R^{exp}$. As it can be seen from Fig. 1a,
$\Delta_R^{th}$ rises very slowly, while $\Delta_R^{exp}$ shows a rapid rise in this
region.

The result of the polynomial fit, with $\chi^2/pt$ value of 0.73, is shown in Fig. 1b.
The corresponding value of $t_{min}^{pp}$ is
 $t_{min}^{pp}=-0.0325 \pm 0.0025\mbox{ GeV}^2 $,
significantly different from the value $t=-0.024\mbox{ GeV}^2$ where
$\Delta_R^{th}(s,t_i)$ goes through zero. We can therefore evaluate, from Eq.
(\ref{ReFN}),
$ReF_N^{pp}(\sqrt{s}=52.8 \mbox{ GeV},\ t=t_{min}^{pp})= 0.375\pm 0.037 \mbox{
(GeV)}^{-2} $
and, from Eq. (\ref{im}), $ImF_N^{pp} (\sqrt{s}=52.8 \mbox{ GeV},\
t=t_{min}^{pp})=7.027  \mbox{(GeV)}^{-2} .$ Therefore $ \rho^{pp} (\sqrt{s}=52.8 \mbox{
GeV},\ t=t_{min}^{pp})=0.053\pm 0.005\ $,
a value which is somewhat different ($\sim 2$ standard deviations) from the value given
in Ref. [5]: 
$ \rho^{pp} (\sqrt{s}=52.8 \mbox{ GeV},\ t=t_{min}^{pp})=0.077\pm 0.009 .$
 The difference in $\rho$-values
  is not highly significant, but it shows the power of our method in the
case of high-precision experimental data.
 The calculation presented here points out toward a real new effect
revealed by our method. This new effect might simply mean that $\rho$ is not a constant
but a function of $t$, as well as $B$ might not be a constant but also a function of
$t$. In others words one must make the analysis of the experimental data with more
sophisticated analytic forms of the scattering amplitude that the exponential one.

Our method uses a given model for $ImF_N^{pp}$ which is supposed to describe well the
experimental data. We know (e.g. from the Regge model) that the forward hadron
scattering amplitude is predominantly imaginary. Therefore a model which describes well
the experimental $dN/dt$ data necessarily has a good $ImF_N(s,t)$ for high $s$ and
small $t$, even if its real part $ReF_N(s,t)$, as a small correction, could be wrong.
In other words, our method is quasi model-independent : different models for
$ImF_N(s,t)$ lead to a quite restricted range of values of $t_{min}$. This is
explicitly shown in Fig.~1c, where we plot $\Delta_R^{exp}(\sqrt{s}=52.8\mbox{
GeV},t_i)$ computed from a model dynamically different from the exponential form, the
Gauron-Leader-Nicolescu (GLN) model \cite{gau88}. This model builds the scattering
amplitudes from the asymptotic theorems constraints as a combination of Bessel
functions and Regge forms, embodies the Heisenberg-Froissart $\ln^2s$ behavior for
$\sigma_T$ and includes the maximal Odderon \cite{luk73}. In this case, $\rho(s,t)$ at
a given $s$ is no more a constant but varies with $t$. This dynamical characteristics
are translated through the fact that $\Delta_R^{GLN}$, as it can be seen from Fig. 1c,
has a \textit{fast increase} in the region $ 0.03<-t<0.06 \mbox{ GeV}^2\ $ , in
agreement with the increase shown by $\Delta_R^{exp}$. The disagreement between
$\Delta_R^{th}$ and $\Delta_R^{exp}$ is seen also through the values of $\chi^2/pt$.
The overall $\chi^2/pt$ value is comparable with the one in the exponential model case:
2.3/pt for a total of 34 points.

We conclude that neither the exponential model nor the GLN model can reproduce entirely
the effect discussed in the present paper : the disagreement between $\Delta_R^{th}$
and $\Delta_R^{exp}$. However, the \textit{stability} of the value $t_{min}^{pp}$
extracted from $\Delta_R^{exp}$ is remarkable: in both models examined in the present
paper this this value is perfectly compatible with the value obtained by polynomial
fit. 

\begin{figure}
 \psfig{figure=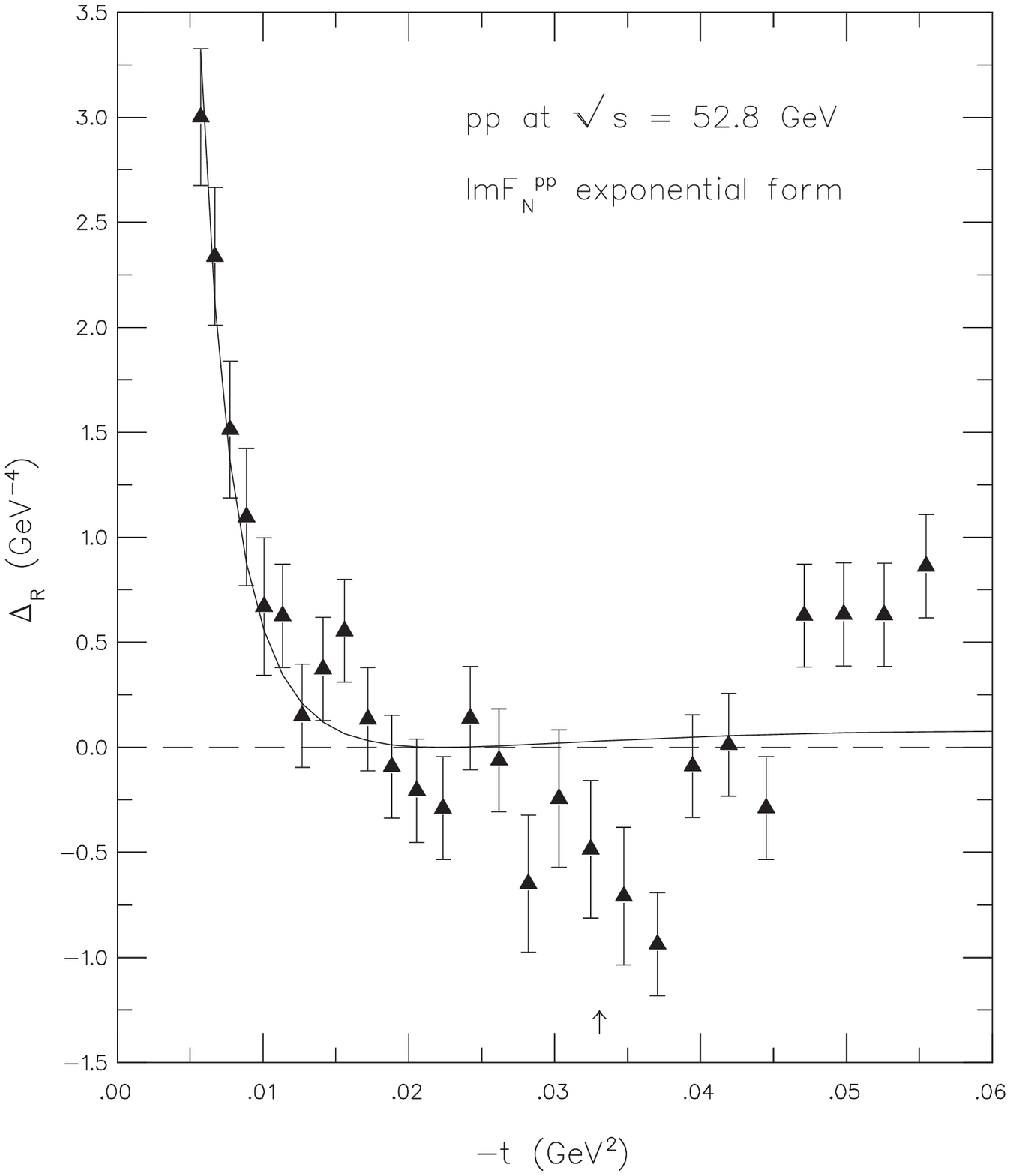,height=2.2in}
\hspace{.2cm}
 \psfig{figure=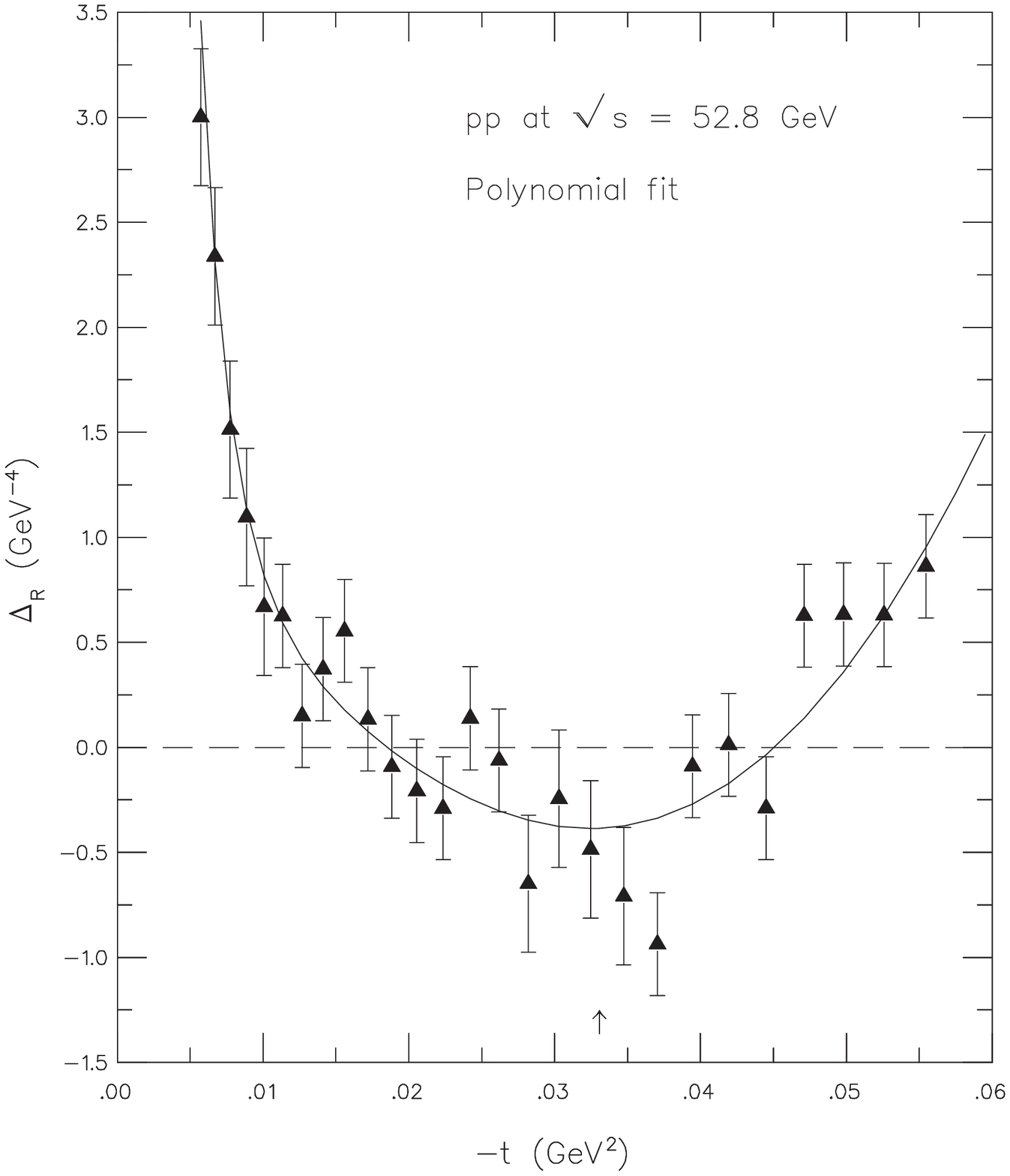,height=2.2in}
\hspace{.1cm}
 \psfig{figure=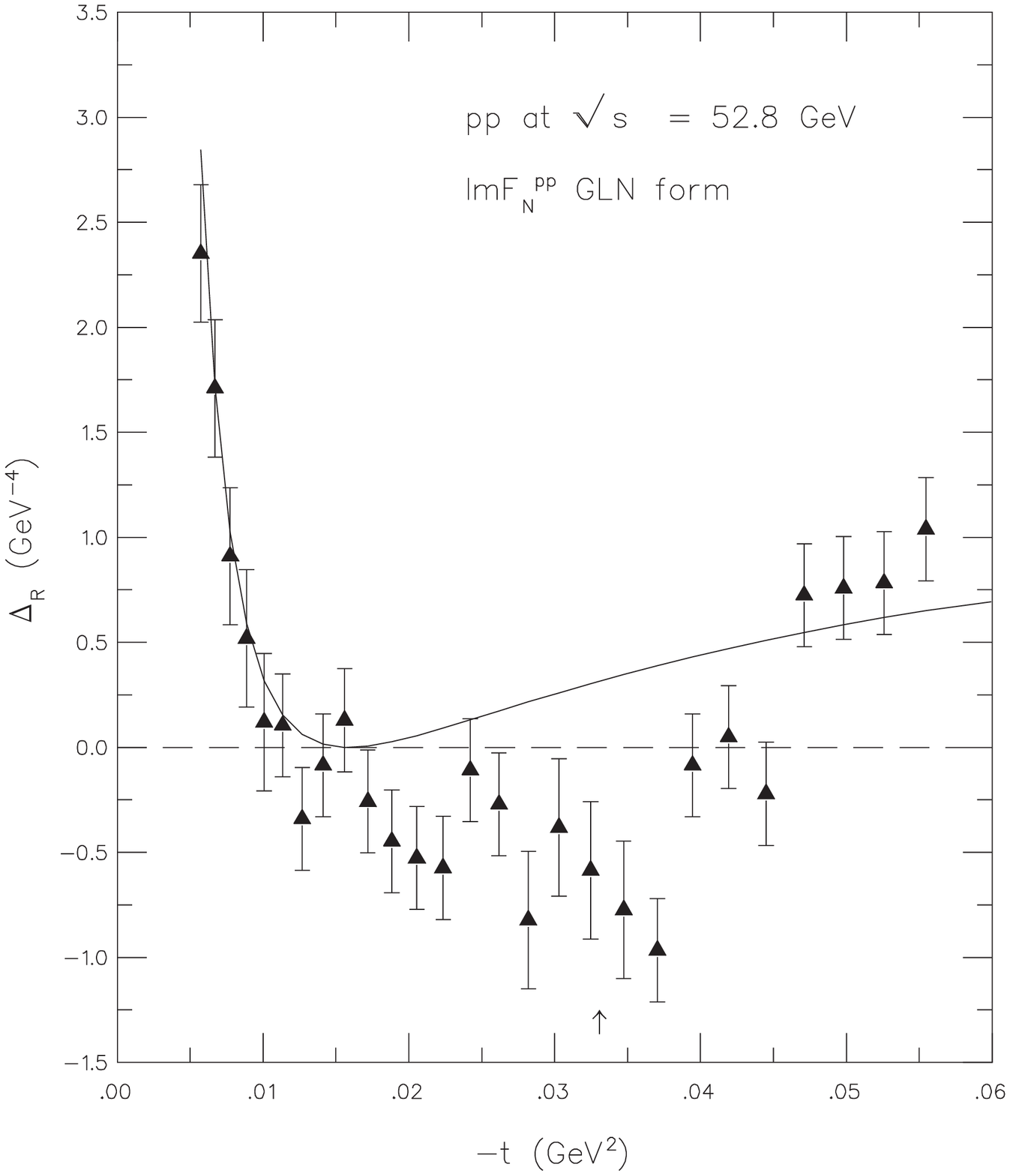,height=2.2in}
\center{ { a)} \hspace{4.5cm} { b)} \hspace{4.5cm} { c)}}
 \caption{ a)
$\Delta_R^{th}$ (the solid curve) and $\Delta_R^{exp}$ (the triangle points) for $pp$
scattering (Eqs. (\ref{Del}) and (\ref{Del2})) at $\sqrt{s}=52.8$ GeV as a function of
$t$, computed with  the exponential form of the amplitude (Eqs. (\ref{im}) and
(\ref{new15})).
  b) $\Delta_R^{exp}$ fitted by the polynomial form
  (the solid curve);
  c)$\Delta_R^{th}$ (the solid curve) and $\Delta_R^{exp}$ (the triangle points)
for  $pp$ scattering (Eqs. (\ref{Del}) and (\ref{Del2})) at $\sqrt{s}=52.8$ GeV  as a
function of $t$, computed within the GLN model
 Ref. [6] }
\end{figure}

In conclusion, we did find a new method for the determination of the real part of the
elastic proton-proton amplitude at high $s$ and small $t$ at a given point
$t_{min}^{pp}$ near $t=0$. The real part of the hadron amplitude is computed, at
$t=t_{min}^{pp}$, from the known Coulomb amplitude. This method provides a powerful
consistency check for the existing models and data and has a predictive power for the
future measurements of the $\rho$-parameter at LHC.

Our method requires high-precision data and a large number of experimental points. We
illustrated how our method works by using the data at $\sqrt{s}=52.8$ GeV (Ref. [5].

As a byproduct of our method we discovered two new effects in the data at
$\sqrt{s}=52.8$ GeV: 1. the significant discrepancy between $\Delta_R^{th}$ as defined
in Eq. (\ref{Del}) and $\Delta_R^{exp}$ as defined in Eq. (\ref{Del2}), $\Delta_R^{th}$
involving $ReF_N$ while $\Delta_R^{exp}$ involves $ImF_N$; 2. $\Delta_R^{exp}$ goes
through a minimum around a $t$-value $\vert t\vert\simeq 0.030-0.035\mbox{ GeV}^{-2}$
and has a sharp increase after this $t$-value.

The dynamical origin of these effects is still obscure. Maybe they are a result of
oscillations in the very small $t$ region. In order to clarify their dynamical origin,
high-precision experimental data at a high energy other than  $\sqrt{s}=52.8$ GeV are
needed. In principle, the experiments which will be performed at LHC \cite{efth} could
explore this problem.

Let us note that our method can be easily extended (with minor changes) to
proton-antiproton scattering, by observing that, in this case, it is the combination
$ ReF_N^{\bar pp}- F_C^{\bar pp} $
which must go through zero at some value $t=t_{min}^{\bar pp}$. The method described in
the present paper could be therefore  used to analyze the UA4 data at $\sqrt{s}=541$
GeV \cite{ua42}, a complex work which will be done and presented in a separate paper.
Of course, in general, one expects that $t_{min}^{pp}\neq t_{min}^{\bar pp}$ at fixed
$s$. Our method could be also extended to the case of proton-nucleus scattering at high
energies.

\section*{Acknowledgments}
The authors express their thanks to C. Bourrely, J.-R.~Cudell, J. Cugnon, W. Guryn,
 J.~Soffer and
 E.~Martynov
for fruitful discussions. One of us (O. S.) thanks Prof. J.~Tran~Thanh~Van
  for the financial support to take the participation on this conference.

\end{document}